\documentclass[prl, amsfonts, twocolumn, nofotinbib, showpacs]{revtex4}
\usepackage{graphicx, epsfig}
\usepackage{color}
\newcommand{\be}{\begin{equation}}
\newcommand{\ee}{\end{equation}}
\newcommand{\bea}{\begin{eqnarray}}
\newcommand{\eea}{\end{eqnarray}}

\newcommand{\gapp}{\mathrel{\raise.3ex\hbox{$>$}\mkern-14mu \lower0.6ex\hbox{$\sim$}}}
\newcommand{\lapp}{\mathrel{\raise.3ex\hbox{$<$}\mkern-14mu \lower0.6ex\hbox{$\sim$}}}
\def\bbox{{\,\lower0.9pt\vbox{\hrule \hbox{\vrule height 0.2 cm
\hskip 0.2 cm \vrule  height 0.2 cm}\hrule}\,}}

\begin{document}
\title{Electroweak stars:  how nature may capitalize on the standard model's ultimate fuel}
\author{De-Chang Dai$^1$, Arthur Lue$^2$, Glenn Starkman$^2$ and Dejan Stojkovic$^1$}
\affiliation{$^1$Department of Physics,
SUNY at Buffalo, Buffalo, NY 14260-1500}
\affiliation{$^2$CERCA, Department of Physics, Case Western
Reserve University, Cleveland, OH~~44106-7079}

 %%%%%%%%%%%%%%%%%%%%%%%%%%%%%%%%%%%%%%%%%%%%%%%%%%%%%%%

\begin{abstract}

\widetext
We study the possible existence of an {\em electroweak star} - a compact stellar-mass object whose central core temperature is higher than the electroweak symmetry restoration temperature. We found a solution to the Tolman-Oppenheimer-Volkoff equations describing such an object. The parameters of such a star are not substantially different from a neutron star - its mass is around $1.3$ Solar masses while its radius is around $8$km. What is different is the existence of a small electroweak core. The source of energy in the core that can at least temporarily balance gravity are standard-model non-perturbative baryon number (B) and lepton number (L) violating processes that allow the chemical potential of $B+L$ to relax to zero. The energy released at the core is enormous, but gravitational redshift and the enhanced neutrino interaction cross section at these energies make the energy release rate moderate at the surface of the star. The lifetime of this new quasi-equilibrium can be more than ten million years. This is long enough to represent a new stage in the evolution of a star if stellar evolution can take it there.
\end{abstract}

%%%%%%%%%%%%%%%%%%%%%%%%%%%%%%%%%%%%%%%%%%%%%%%%%%

\maketitle

\section{ Introduction}
The last stage of the evolution of a star massive enough to go through the supernova explosion (but not massive enough to become a black hole) is believed to be a neutron star. A neutron star has no active energy source which balances the force of gravity. Instead, it is supported by degeneracy pressure, which arises due to the Pauli exclusion principle. However,  this pressure can support only neutron stars lighter than about $2.1 M_\odot$, which is known as the Tolman-Oppenheimer-Volkoff  limit. Neutron stars heavier than this limit eventually become black holes.

Recent studies point out the existence of a state between the neutron star and black hole. It is called a quark star. This state owes its existence to the QCD phase transition in which the original nuclear matter becomes quark matter. This process can release $10^{53}$ erg in about $10^{-3}$ to $10^{-2}$s \cite{Drago:2005yj,Drago:2008tb} in the form of neutrino bursts. While this a huge amount of energy, it is released in too short a time to provide the pressure to prevent gravitational collapse. After this energy release, a star containing effectively only three quark flavors (u,d,s) can exist in a stable equilibrium where the pressure is provided by the Pauli exclusion principle. However, at higher densities where four or more quark flavors are present, quark matter cannot avoid gravitational collapse.

Since in the gravitational collapse matter gets compressed to ever increasing densities/temperatures, it is natural to explore what could happen at the electroweak phase transition, the next (and last) within the standard model.
In the standard model, both baryon and lepton number $U(1)$ global symmetries are accidental, conserved
perturbatively, but violated by non-perturbative processes such as those mediated by instantons. At temperatures
well below the electroweak symmetry-breaking scale, $T\lapp100$GeV, these non-perturbative processes are suppressed by the extremely small factor $e^{-8\pi/\alpha}$ (giving, for example, a proton life-time of $10^{141}$yrs \cite{Adams:1996xe}). (Note, unless otherwise indicated we shall employ natural units where $\hbar=c=k_B\equiv1$.)
However, above this scale, baryon and lepton number  violating processes are expected to be  essentially
unsuppressed (although the combination $B-L$ remains conserved).
Quarks can then be effectively converted into  leptons. In this process, which we call {\em electroweak burning}, huge amounts of energy can be released.

Let us assume that in the core the temperature is above, or at least very near, the
weak scale, and the matter density is also comparable $\rho\gapp (100{\mathrm GeV})^4$.
At these densities, the matter is opaque even to neutrinos  and the energy released at the center cannot stream freely out of
the star. Nevertheless, the energy can be carried out, mostly by neutrinos and anti-neutrinos.  These can also carry out any excess anti-lepton number generated in the electroweak burning, and thus prevent the lepton number chemical potential from halting the consumption of the baryon number.
This mechanism can likely provide a stable energy source which can counteract gravity for a while.
On a side track, we remark that studying the electroweak star physics,
in particular the consequences of a very powerful energy source that was  so far neglected,
may provide important  new clues about supernova explosions, but we reserve further discussion on
this matter to a future paper (in preparation).

In this paper, we study the inner structure of such a star and calculate its lifetime. We find that this new phase can last up to $10^{15}$s, which is long enough to give a new name to this type of stars - electroweak stars.  In order to find out whether electroweak stars are very common or very exotic objects, much more detailed analysis is required.  Future work will consider questions of stability, of evolution  -- if and when the electroweak star arises in the late-phases of  evolution of ordinary stars -- of the structure of the outer cooler layers, and of observational signatures  of these objects.

\section{ Density, pressure, and particle number density inside the electroweak star}
\begin{figure}[t]
   \centering
\includegraphics[width=3.5in]{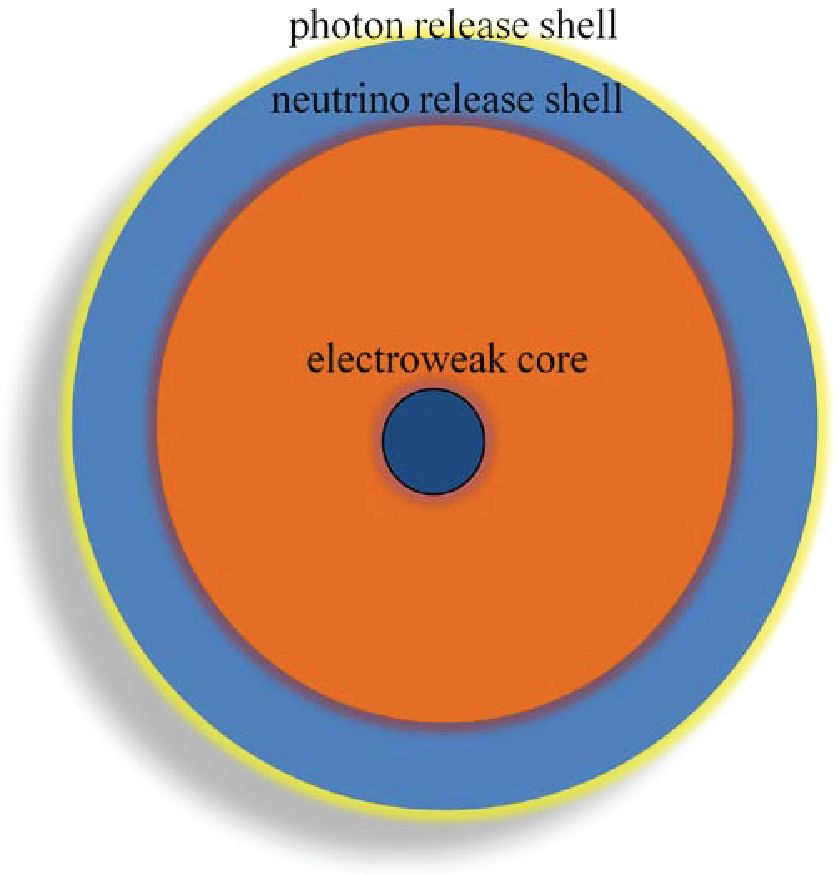}
\caption{The structure of an electroweak star. We separate the core of the star into three regions: I. central core where the electroweak symmetry is restored, II. region where the high energy neutrinos are trapped, III. region from which high energy neutrinos can escape. Neutrinos will be emitted between the regions II. and III. Photons are emitted at the surface of the star.}
    \label{fig:structure}
\end{figure}

We separate the core of the electroweak star into three regions, as shown in Fig.~ \ref{fig:structure}. The central region is hotter and denser  than the electroweak symmetry restoration temperature  ($T\gapp100$GeV).
This region is very dense ($\rho >10^{8}$GeV$^4$), but small (several cm).
We find that the total mass inside this region is $\sim 5\times10^{-6} M_{\odot}$.
Since the lepton and baryon number are not conserved in this region, but $B-L$ is,
baryons are freely transformed to anti-leptons so long as it is thermodynamically favorable to do so.
A $SU(2)$-preserving instanton interaction can convert $9$ quarks into $3$ anti-leptons \cite{Manton:1983nd},
 for example:
\be
\label{eqn:instanton}
udd \ \  css \ \  tbb \ \  \rightarrow  {\bar \nu_e} , {\bar \nu_\mu} , {\bar \nu_\tau}\ee
and all $B-L$ and electromagnetic charge conserving  variations involving quarks, anti-quarks, neutrinos, antineutrinos
and charged leptons. At these densities and temperatures each particle carries about $100$GeV of energy, so this process can release about $300$GeV per neutrino. The typical energy of the produced neutrino is not the only relevant quantity here, we also need the rate at which the energy is released. The sphaleron baryogenesis/baryodestruction rate (above the electroweak symmetry restoration temperature) is proportional to $T^4$. However, we will show later that the energy release rate from the core is limited by the quark supply rate (dictated by the gravitational collapse) rather than the maximal electroweak burning rate.

The energy in the form of neutrinos  flows out of the central core, and eventually out of the star.  However, in the central region,
the mean free paths of all particles are short compared to the size of the star. The energy must therefore diffuse out of
the central core.  Similarly, the non-perturbative B-violating processes, while reducing B,  reduce L.  Initially
$B-L>0$  in stars, because neutrinos escape, and thus there are more neutrons plus protons than there are electrons.
Thus, the B-violating processes would cause an anti-lepton number to build up.  This must either stop the burning, or be carried out of the core by a flux of anti-leptons.
Outside the central core, baryon and lepton number are conserved.   The density falls with increasing radius,
and particle mean free paths increase,  however, they are still too short for  particles to stream freely out of the star.
Eventually, however we reach the star's neutrino release shell (i.e. neutrinosphere), where the neutrino (and anti-neutrino)
mean free path is long enough for these particles to flow freely. [This is well within the radius at which the
same is true of photons.]   The neutrinos and anti-neutrinos carry off not just most of the energy produced in the core
burning region  (the rest is carried off by photons),
but also the lepton number, thus permitting the burning to continue.  Since the fuel is predominantly
baryons not anti-baryons, resulting in the production of an excess of anti-leptons over leptons, there will be a
greater flux of anti-neutrinos than neutrinos from the star.
Moreover, the electroweak processes (unlike inverse beta decays in normal stars)
creates equal numbers of anti-neutrinos of all three generations (see Eq.~\ref{eqn:instanton}).
Thus the excess is likely to be equal in all generations.
Such a signal, taking into account neutrino oscillations, might be an indication that the electroweak processes are taking place at the core of the star.
However, the excess  is likely to be very small,  since (as in a supernova),
the emission is expected to be dominated by thermal emission from
the neutrinosphere (which carries no excess),
not by direct anti-neutrino production in the core.
More likely, a detailed analysis of the electroweak star will show that the neutrinospheres for the
three generations of neutrinos and anti-neutrinos are separated due to the differing masses
and hence abundances and momentum distributions as a function of radius of the quarks and
charged leptons of the associated generations.  Therefore the fluxes and spectra of the
different neutrino and anti-neutrino species should be predictably (and hopefully observably)
different.

The size of a quark star is about $10$km, while its mass is of the order of $M_\odot$. A small electroweak core cannot significantly change the region outside the core.
We use the so-called ``bag model"
\cite{Chodos:1974je} as an approximate solution of the region outside the core.
The pressure and energy density are \cite{Kettner:1994zs}
\begin{eqnarray} \label{pe}
P&=&\sum_{i} p_i -B_N\\
\epsilon &=&\sum_{i} \epsilon_i +B_N. \nonumber
\end{eqnarray}
Here, $P$ is the total pressure, $p_{i}$ is the partial pressure of particle $i$, $B_N$ is the bag energy density, $\epsilon$ is the total energy density, and $\epsilon_i$ is the energy density of  particle $i$. The subscript $i$ runs over all types of particles present in the star (however here for simplicity we include only fermions, i.e. quarks, neutrinos and electrons). We choose the specific value $B_N^{1/4}=145$MeV. The condition for the electric charge neutrality is
 \be \label{charge}
 \sum_i q_i n_i =0.
 \ee
Here, $q_i$ is the charge of the particle $i$. The pressure, energy density, and number density of particles can be
well approximated from Fermi-Dirac and Bose-Einstein gas distribution:
\begin{eqnarray} \label{pne}
p_i&=&\frac{g_i}{6\pi^2}\int^\infty_{m_i}dE(E^2-m_i^2)^{3/2}f_i(E)\\
\epsilon_i &=& \frac{g_i}{2\pi^2}\int^\infty_{m_i}dE (E^2-m_i^2)^{1/2}E^2 f_i(E) \nonumber \\
n_i&=&\frac{g_i}{2\pi^2}\int^\infty_{m_i}dE (E^2-m_i^2)^{1/2}E f_i(E) \nonumber.
\end{eqnarray}
Here, $g_i$ is the number of degrees of freedom of particle $i$ (e.g. $g=2$ for leptons and $g=6$ for quarks), $m_i$ is the mass of the particle, $\mu_i$ is the chemical potential of the particle and $T$ is the temperature.  The distribution functions are
 \be \label{fi}
 f_i(E)=\frac{1}{1\pm e^{(E-\mu_i)/T}}
 \ee
 for bosons and fermions.

\section{Solution to the Tolman-Oppenheimer-Volkoff equations describing an electroweak star}

In this section we setup and numerically solve the system of equations that describe the structure of an electroweak star.

Outside the central region I, baryon and lepton numbers are conserved.
We assume for simplicity that the chemical potentials of the left- and right-handed fermions are the same,
and that all fermions of the same charge (e.g. $u$, $c$ and $t$ quarks) have the same chemical potential.
An antiparticle has the chemical potential of the opposite sign. The interaction
$d \rightarrow  u+e+\bar{\nu}_{e}$ gives one relationship
\be \label{mu1}
 \mu_d=\mu_u+\mu_e-\mu_\nu.
 \ee
 Eq.~(\ref{eqn:instanton}) implies another relation inside the electroweak core:
 \be \label{mu2}
 \mu_u +2\mu_d+\mu_\nu=0,
 \ee
We assume $\mu_\nu=0$ outside the electroweak core, so that neutrinos or anti-neutrinos can propagate away.
We set the boundary conditions at the center that pressure and energy density are  $(100$GeV$)^{4}$, and then calculate the structure of the star according to the Tolman-Oppenheimer-Volkoff equations
\begin{eqnarray} \label{structure}
\frac{dP}{dr}&=&-\frac{(\epsilon +P)(M+4\pi Pr^3)}{M_p^2 r^2 (1-\frac{2M}{M_p^2 r})},\\
\frac{dM}{dr}&=&4\pi \epsilon r^2. \label{M}\\
\label{eq:gtt}
g_{tt}&=&(1-\frac{2M_{\rm star}}{M_p^2 R_{\rm surface}})e^{-\int^{P(r)}_0 \frac{2dP}{P+\epsilon}}\\
g_{rr}&=& \frac{1}{1-\frac{2M(r)}{M_p^2 r}}  \label{eq:grr}
\end{eqnarray}
$R_{\rm surface}$ is radius of the star, $M_{\rm star}$ is the mass of the star, while $M_p$ is the Planck mass. Since we are dealing with very dense objects, general relativistic effects described by the metric components $g_{tt}$ and $g_{rr}$ will be very important.

If we supply an equation of state $\epsilon(P)$, the Tolman-Oppenheimer-Volkoff equations could be solved directly. However, we do not want to assume an equation of state, and instead we try to solve the system of coupled equations we already set-up.

Because of the properties of Fermi-Dirac statistics, fermions can have large energies at $T=0$. In that case the energy of neutrinos is not proportional to the temperature, instead it is proportional to the chemical potential $\mu_\nu$, as can be seen from Eq.~(\ref{fi}). We will use this fact to eliminate $T$ from the equations and simplify our calculations.
The exact treatment would require keeping $T$ as a variable (beside the energy density, pressure and number density), setting the boundary condition at the center $T=100$GeV, and finding the temperature profile $T(r)$. Setting $T=0$ will not significantly change $P(r)$, $\epsilon (r)$ and $M(r)$, but may change $n(r)$. This change in $n(r)$ will seemingly affect the neutrino mean free path, but one can explicitly verify that this does not happen since the mean free path in Eq.~(\ref{eqn:mean_free}) will effectively depend only on  the energy density $\epsilon_i = n_i E_i$. Thus, the structure of the star will not change much if we set $T=0$. [Note that the properties of Bose-Einstein statistics are different and one cannot set $T=0$ when including photons and weak gauge bosons into consideration.]

Thus, we have four independent equations (\ref{charge}), (\ref{mu1}), (\ref{mu2}), and (\ref{structure}), with four independent variables $\mu_u$, $\mu_d$, $\mu_e$ and $\mu_\nu$. The chemical potentials $\mu_u$, $\mu_d$, $\mu_e$ and $\mu_\nu$ then uniquely specify $P(r)$, $\epsilon (r)$, $n_i(r)$ and $M(r)$ through equations (\ref{pe}), (\ref{pne}), (\ref{fi}), (\ref{structure}) and (\ref{M}).

\begin{figure}[t]
   \centering
\includegraphics[width=3in]{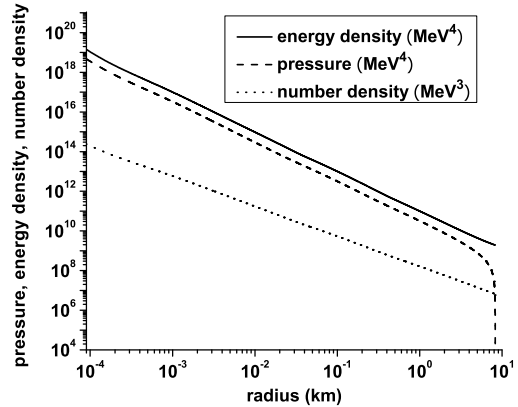}
\caption{The pressure, energy density and particle number density change with the radius of the star. The total radius of the star is about $8.2$km (radius where pressure and energy density drop to zero), while  the mass is about $1.3M_\odot$ (mass within that radius).}
    \label{fig:density}
\end{figure}

The numerical results are plotted in Fig.~\ref{fig:density}.
The total size of the star is $8.2$km (radius where pressure and energy density drop to zero) and its mass is $1.3 M_\odot$ (mass within that radius).
The size of the central electroweak core is several cm (this is the distance at which the density falls below $(100$GeV$)^{4}$. Within this central region, the pressure and energy density gradients are very small, so pressure and energy density are practically constant there.
We can also find the dependence $\epsilon (P)$ and find that the equation of state does not change much across the star and is approximately $\epsilon =3 P$. In the absence of the active energy source, such an equation of state would signal instability. However, electroweak stars have a very powerful energy source which can balance gravity. Also, this equation of state will tip toward a non-relativistic equation of state once weak gauge bosons (and interactions among the particles) are included into calculation, since they are not ultra-relativistic at these energies.

\begin{figure}[t]
   \centering
\includegraphics[width=3in]{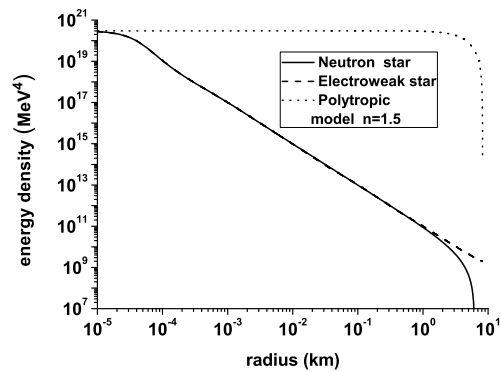}
\caption{Comparison between the energy density profiles of an electroweak star, neutron star (a star made from neutrons only) and polytropic star (with the polytropic index $1.5$). For all three plots the central density is $3\times 10^{20} MeV^4$. The neutron and electroweak stars are almost indistinguishable at small radii because particle masses are not important in that region, but they become distinguishable at large radii.}
    \label{fig:density:comp}
\end{figure}

For comparison, in Fig.~\ref{fig:density:comp} we plot the energy density profiles of an electroweak star, a neutron star (a star made from neutrons only) and a polytropic star (with the polytropic index $1.5$), where, to emphasize the difference (and similarity) in the structures, we set the  central density for all three  to $3\times 10^{20} MeV^4$. (Of course, no stable neutron star can have this high a central density.) The neutron and electroweak stars are almost indistinguishable at small radii because particle masses are not important in that region, but they become distinguishable at large radii.  Furthermore, we are currently treating the quarks in the electroweak star as a degenerate gas of non-interacting particles.  Adding in the interactions, which is likely to improve the stability of the electroweak star, is the subject of a future work.

The non-perturbative electroweak interactions provide the source of energy that keeps the star from collapsing.
As the baryons are burned,
gravitational collapse  brings  more material to the central part of the core feeding the fire.
We assume that eventually a quasi-equilibrium between the burning and the refueling is achieved.
If so, the star enters a quasi-stable state that can last for a while. The question of dynamical stability will be
addressed in future work.

\section{ Location of the neutrinosphere}
Neutrinos near the central region have very high energies (but below the weak scale)
and the local density is extremely high.
Their mean free path in this region is thus very short, and they cannot stream freely from the star.
However, as the local density falls with radius, and the neutrinos' average energy decreases,
their mean free path increases.  We define  the neutrino escape radius (as a function of neutrino energy)
as the distance from the center at which a  neutrino with  that energy has a mean free path
greater than the thickness of the overlying matter and can therefore escape the star.
The neutrinosphere is the set of such spherical shells for neutrinos of energies characteristic
of the local temperature.

Now we study the neutrino mean free path in dense media as presented in \cite{Shen:2003ih}. We assume that the cross section of neutrino scattering is \cite{Marciano:2003eq}
\begin{equation}
\label{eqn:cross}
\sigma_i \sim \frac{G_F^2 s}{\pi}\sim \frac{G_F^2 E_\nu E_i}{\pi}.
\end{equation}
Here, $G_F\sim 1.166\times 10^{-5}$GeV$^{-2}$ is Fermi constant, $s$ is the center of mass energy, $E_i$ is the energy of the particle $i$, while $E_\nu$ is the neutrino energy. The mean free path is
\begin{equation}
\label{eqn:mean_free}
\frac{1}{\lambda}= \sum_i \sigma_i n_i
\end{equation}

\begin{figure}[t]
   \centering
\includegraphics[width=3in]{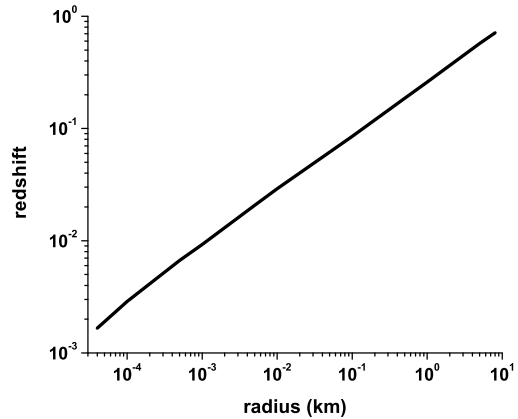}
\caption{The redshift factor $\sqrt{g_{tt}(r)}$ vs. radius of the star. The redshift near the center is much larger than at the surface. A particle with the original energy of $100$GeV near the center carries away only $165$MeV as it leaves the surface. Opacity of the dense medium will further reduce this energy.}
    \label{fig:redshift}
\end{figure}

\begin{figure}[t]
   \centering
\includegraphics[width=3in]{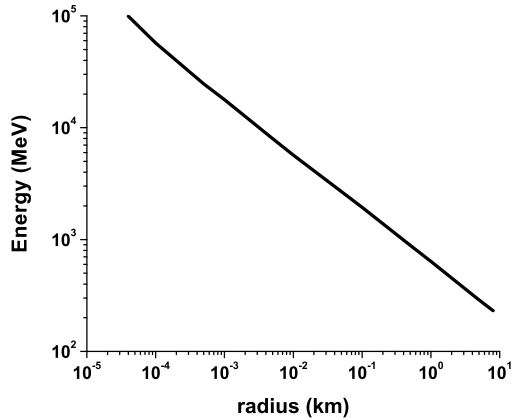}
\caption{Neutrino energy as a function of the distance from the center where only effects of gravitational redshift are taken into account (no energy loss due to interactions). The energy on the surface is about $1000$ times smaller than the original energy at the center.}
    \label{fig:nu_energy}
\end{figure}

Unlike any other active star, particles propagating through the electroweak stars suffer large gravitational redshift. Thus $E_\nu$ is not a constant, but changes as
\begin{equation}
\label{eqn:E_nu}
E_\nu (r)=\frac{\sqrt{g_{tt}(r_0)}}{\sqrt{g_{tt}}(r)}E_\nu (r_0).
\end{equation}
The redshift ratio is shown in Fig~\ref{fig:redshift}. Obviously, the redshift effect near the center is much larger than at the surface. Since inside the star $\epsilon \sim 3P$, the redshift can be estimated from Eq.~(\ref{eq:gtt}). It is
\begin{equation}
\label{eq:decate}
E_\nu \sim \left(\frac{P(r)}{P(r_0)} \right)^{1/4} E_\nu ({r_0})
\end{equation}
The pressure is about $(100$GeV$)^{4}$ near the center, and about $(100$MeV$)^{4}$ near the surface
(before the bag energy changes the pressure a lot).
Therefore, a particle retains only a fraction  $10^{-3}$ of its original energy.
Though instanton processes release huge amounts of energy at the center,
the star can emit much more moderate amounts.
The result is shown in Fig.~\ref{fig:nu_energy}. A particle with the original energy of $100$GeV near the center indeed carries only
$165$MeV as it leaves the surface due to the redshift. This energy will be further reduced due to opacity of the high density medium that neutrinos are traveling through.

We now use this energy corrected for the redshift to calculate the mean free path on the surface.
From Eq.~(\ref{eqn:cross}) and Eq.~(\ref{eqn:mean_free})
the mean free path on the surface is
\begin{equation} \label{mfp}
\frac{1}{\lambda} \sim \frac{G_F^2 \epsilon}{\pi} E_\nu (R_{\rm surface})
\end{equation}
Eq.~(\ref{mfp}) yields  $\lambda \sim 10^{12}$MeV$^{-1}\sim 10^{-4}$km. At the core, Eq.~(\ref{mfp}) yields  $\lambda \sim 10^{-14}$m. A neutrino with this energy cannot leave the star freely, and must lose its energy while propagating from the center to the surface. This energy loss will in turn further change the neutrino mean free path. Properly calculated  neutrino energy (plotted in Fig.~\ref{fig:energy-core}) will imply that the neutrino mean free path at the neutrinosphere is $\lambda \sim 0.1$km.

With the redshift taken into account, the neutrinos have roughly the same energy as the background they are propagating in, which is true in general for the other types of particles. If a particle has the energy density $\epsilon$, its energy can be estimated as $\sim \epsilon^{1/4}\sim P^{1/4}$.
Comparing this with Eq.~(\ref{eq:decate}), the energy of a neutrino decreases with the distance the same way as the background. Thus,
the neutrinos coming from the center and arriving to a certain point inside the star have roughly the same energy as the neutrinos that were emitted by thermal processes at that point.

\section{ Transport of energy through the star, energy release rate, and the lifetime of an electroweak star}

We would like now to  estimate the (neutrino) luminosity of the electroweak star.
A naive luminosity (i.e. energy release rate) can be obtained from the black-body radiation
at the stars core
\be \label{err}
{\cal L}_\nu \leq {\cal L}_{bb} \sim \sigma_{bb} \epsilon_\nu \pi R_{\rm core}^2
\ee
This places an upper bound of about  $3\times 10^{41}{\mathrm MeV}^{2}=7\times10^{56}{\mathrm erg/s}$.
At this rate it would take less than a second to release $1M_\odot$.
Fortunately, Eq.~(\ref{err}) is a severe over-estimate of the luminosity.
It does not take general relativity into account,
nor does it allow for the fact that the luminosity depends not just on the temperature and density at the neutrino-sphere
but also on their gradients, since those determine the {\em net} outward flux of energy.

First we will estimate the energy release rate when we include redshift effects in calculating the ratio $dE/dt$.
An upper bound on the energy release rate can be obtained from the free fall time of the incoming quark shell into the electroweak-burning core.
This determines the maximum rate at which the electroweak core burning can be fed.
The free fall acceleration over a short distance given by the mean free path $\lambda$ is roughly given by $a=M_{\rm ew}/r_{\rm ew}^2$, where $r_{\rm ew}$ refers to a radius one mean free path ($\lambda$) away from the electroweak core (relativistic corrections will not change the result significantly). Therefore, the energy release rate can be no more than\begin{equation} \label{td}
\left( \frac{dE}{dt} \right)_{\rm max} \sim 4\pi r^2_{\rm ew} \sqrt{ \frac{2\lambda M(r_{\rm ew})}{r^2_{\rm ew}}}\, \epsilon (r_{\rm ew}) \, g_{tt}(r_{\rm ew}),
\end{equation}
This is just  $10^{27}$MeV$^2$, which is already much lower than the earlier upper limits obtained from Eq.~(\ref{err}).
We compare this rate with the electroweak baryogenesis/baryodestruction interaction rate, which is above the electroweak phase transition proportional to $T^4$ \cite{Moore:1998zk}
\begin{equation}
\frac{dE_w}{dt}\sim 4\pi r^2_{\rm ew} \times 20 \alpha_{\rm ew}^5 T^4
\end{equation}
This rate is about $10^{34}$MeV$^2$, which is much larger than the rate in (\ref{td}).
Therefore we can reasonably assume that as quarks reach the electroweak core,
they are converted into neutrinos instantaneously.
Otherwise, the in-falling  matter would pile up and form a black hole. This result also implies that the star's structure and energy released at the surface do not crucially depend on precise details of the electroweak processed at the core, i.e. the star is capable of burning much more quarks than it actually does.

We now calculate the neutrino energy release rate at infinity more accurately taking into account the relativistic transport of energy through the star. Gravitational redshift and time delay along with the enhanced neutrino interaction cross section at these energies will affect the energy transport through the star and make the energy release rate moderate at the surface of the star. Imagine two surfaces inside the star whose radii are one $\lambda$ apart. The outward flux of neutrinos will be partially canceled by the backward flux due to neutrino collisions with other particles, and we are interested only in the net outward flux. The relativistic transport of neutrino energy can then be described by
\begin{equation}
4\pi r^2\lambda\frac{d ( S (r)g_{tt}(r))}{\sqrt{-g_{rr}(r)}dr}=-L.
\end{equation}
Here $S(r) \equiv dE(r)/(dt dA)$ is the energy flux at the radius $r$,  $A$ is the area, while $L \equiv dE/dt$ is the energy release rate at infinity (which is about the same order as the intensity at the neutrinosphere).  The factor $g_{tt}(r)$ describes both the redshift and time delay. The factor $g_{rr}(r)$  takes into account that the space-time inside the star is not flat. The energy flux $S(r)$ can be found from the energy density $\epsilon_\nu$ given by (\ref{pne}) knowing the energy (i.e. chemical potential) of neutrinos $E_\nu$. The metric coefficients are given by Eq.~(\ref{eq:gtt}) and (\ref{eq:grr}). It is then straightforward to calculate $L$, given all the other parameters.

The neutrino mean free path is not a constant throughout the star, but changes from point to point.  We therefore set a practical definition of the location of the neutrinosphere as $R_{\rm nsph}$ with
\be \label{rnsph}
\int_{R_{\rm nsph}}^{R_{\rm star}}  \frac{dr}{\lambda(r)} =1 ,
\ee
where $\lambda(r)$ is given by Eq.~(\ref{mfp}). This definition requires that a neutrino, at a given radius $r$ and a given energy $E_\nu$, interacts only once with other particles  before it leaves the star.

\begin{figure}[t]
   \centering
\includegraphics[width=3in]{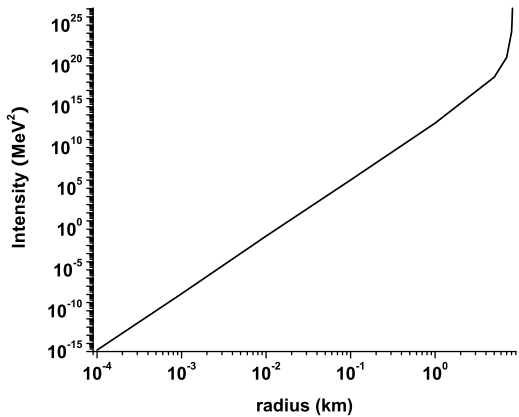}
\caption{The energy release rate vs. neutrino release radius: The energy release rate increases with the radius of the neutrino release shell (i.e. neutrinosphere). The maximum energy release rate is on the surface, but there it exceeds the limit from the quark shell free fall into the electroweak core  (which minimizes the denominator in (\ref{err})). This implies that the neutrino release shell must be inside the star.}
    \label{fig:intensity}
\end{figure}

With these definitions, we can calculate intensity $L$ as the function of the radius of the neutrinosphere $R_{\rm nsph}$. From the condition (\ref{rnsph}), we find the neutrino energy $E_\nu$, which in turn gives the required energy flux $S(r)$ for a given $R_{\rm nsph}$. The result is plotted in Fig.~\ref{fig:intensity}.
Larger radius $R_{\rm nsph}$ implies lower energy density, which means that higher energy neutrinos can escape.
The energy release rate therefore increases with the radius of the shell and it is maximal at the surface of the star. However, there  it exceeds the limit from the quark free fall. This implies that the neutrino release shell must be inside the star.

\begin{figure}[t]
   \centering
\includegraphics[width=3in]{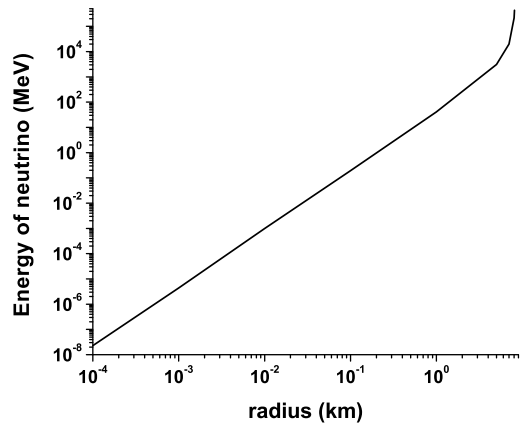}
\caption{The neutrino energy at the surface of the central electroweak core vs. the radius of the neutrino release shell (i.e. neutrinosphere) $R_{\rm nsph}$. From Fig.~\ref{fig:intensity} we see that the energy release rate increases with the $R_{\rm nsph}$. Higher energy release rate needs the source of higher energy. The energy of neutrinos produced at the core therefore increases with the radius of release shell. However, the energy is already larger than $500$GeV at $8.2$km. The star cannot support this amount of energy implying that its radius must be smaller than that. In fact, $300$GeV neutrinos produced by baryon number violating processes at the core imply that the radius of the neutrino release shell is $8.1$km.}
    \label{fig:energy-core}
\end{figure}

At the neutrinosphere, $R_{\rm nsph}$, the following must be true, $4\pi r^2 S(r=R_{\rm nsph})g_{tt}(r=R_{\rm nsph})=L$, since there is no backward flux of neutrinos (all the neutrinos that reach $R_{\rm nsph}$ leave the star freely). From there we can find the neutrino energy, $E_\nu$, at the central electroweak core needed to support a certain intensity at the surface of the star.
Fig.~\ref{fig:energy-core} shows the results. Higher energy release rate needs the source of higher energy.  If the neutrino release radius is $8.1$km, then the original neutrino energy must be about $300$GeV, which is equal to the energy per neutrino that typical electroweak processes can release at the center. We therefore conclude that the radius of the neutrino release shell can be at most $8.1$km.
If the neutrino escape radius is about this size, the energy release rate is about $10^{24}$MeV$^2$. Integrating this, we find that it takes about $10^{15}$sec ($10$ million years) to release  energy equivalent of $1M_\odot$. This is the minimal life-time of the electroweak star, provided that the electroweak burning does not stop before all the quark fuel is spent.

In our calculations we neglected the fact that some fraction of energy is carried away by photons, since they have shorter mean free path. Further, though the net lepton number is conserved in this region, the number of neutrinos and antineutrinos are not conserved.
We also ignored the effects of energy transport due to convection.
Finally, neutrino energy was calculated from the chemical potential change, instead of the temperature change.
However, we assume that these effects will not change the order of magnitude estimate.

On a completely different side, it is not difficult to imagine that the electroweak physics could be responsible for some of {\bf the Type II supernova explosions}.
Many decades have elapsed since modeling of the core collapse supernova mechanism began in detail. However, the exact mechanism remains elusive so far. It appears that the explosion of supernova is a multiscale and multiphysics event, where gravity, neutrino transport, fluid instabilities, rotation, magnetic fields etc are all very important ingredients. The existing models failed to produce the supernova explosion. It will be of high importance to check if the electroweak physics can help supernovae explode by providing a very powerful energy source that was neglected so far in the studies of this problem. The expected final outcome of the stability analysis of the electroweak stars sequence will perhaps yield a stability curve. The possible outcomes are that the whole electroweak sequence is stable or unstable. However, a generic outcome with different regions in parameter space corresponding to stable and unstable stars depending on the initial conditions is more likely. In unstable regions, the electroweak stars with a certain initial mass and central density could be unstable to explosions. These explosions could be in the form of gamma ray bursts or perhaps supernova explosions. To find out the answer, a very detailed modeling of the electroweak star is needed taking all the necessary ingredients like gravity (with full GR effects), neutrino transport, fluid instabilities etc into account.

\section{Conclusions}
We studied the possibility of the existence of a new phase in the stelar evolution. After all of the thermonuclear fuel is spent, and possibly after the supernova explosion, but before the remaining mass crosses its own Schwarzschild radius, the temperature of the central core may surpass the electroweak symmetry restoration temperature. At this point, non-perturbative baryon-number violating interactions in the ordinary $SU(2)\times U(1)$  electroweak standard model become unsuppressed.  The consequent relaxation of the stellar baryon number chemical potential, i.e. the conversion of baryons to anti-leptons, may provide a source of pressure which can balance gravity.
We constructed a solution to TOV equation whose central pressure is non-singular (and thus not a black hole). We showed that the in-falling matter gets converted into neutrinos at the rate much faster than the free-fall rate, which indicates that the mater burns before crossing its own Schwarzschild radius (assuming that the core is not within the Schwarzschild radius itself).  Our analysis shows that lifetime of this new phase can be as long as $10$ million years which we propose to call
an electroweak star.  We emphasize that the electroweak star relies only on the Standard Model physics
for its existence.

Electroweak stars would be an exciting addition to the diverse menagerie of astrophysical bodies that the universe
provides (see also recent proposals in \cite{Spolyar:2007qv}).  Nevertheless, considerable work remains to be done before we can claim with confidence that such objects will form in the natural process of stellar evolution, or that they will indeed burn steadily for an extended
period.  Similarly, assessing the visibility of these fascinating new objects requires a careful modeling of their outer structure to determine
the neutrino/antineutrino \cite{Horiuchi:2008jz} and photon luminosity and spectrum.

While we constructed a non-singular solution to TOV equations which describes the electroweak star, the crucial next step would be to verify that a full time dependent evolution will take the collapsing system to this configuration. This is not unlikely to happen since the global parameters of the electroweak star are not much different from those of the neutron star. The radius is $8.2$km and the mass is $1.3$ Solar masses. What is different is the central core which is at or above the electroweak density. In gravitational collapse, the collapsing object is not crossing its own Schwarzschild radius at once as a whole. The simulations show that the central density is much higher than the rest of the star (since the outer layers push the inner layers stronger and stronger). It appears that it is not difficult to achieve central density of the order of the electroweak phase transition. For example, in \cite{Balberg:2002ue}, the authors find that densities achieved in gravitational collapse are much larger than what would be expected from the selfsimilar solutions (homologous collapse, PPN and other commonly used approximations fail in this regime). In particular, the Fig. 6 in \cite{Balberg:2002ue} shows that, with appropriate initial conditions (say those of a neutron star), the electroweak densities can be reached before the "gravothermal catastrophe", i.e. black hole formation. Once that stage is reached, electroweak burning will get ignited and the collapse will be stopped for a while. This certainly does not imply that the whole star is at the electroweak density, in contrary, it is not much different from the ordinary neutron star, as the TOV solution demonstrates.

One intriguing possibility is that the authors of \cite{Barcelo:2007yk} and \cite{Visser:2009pw} are correct that
gravitational collapse of a stellar mass does not result in a black hole but in an object of very high density and temperature
toward its center. Then one would not have to worry about the collapsing object crossing its own Schwarzschild radius before reaching electroweak densities.

Finally, electroweak phase transition may be triggered inside a neutron star by a mechanism reminiscent of sonoluminescence, where non-linear waves (e.g. star-quakes) may concentrate huge energy within some small volume \cite{Simmons:1996ks}.

We emphasize further that electroweak stars do not have to be formed
exclusively in stellar matter collapse. Primordial black holes are formed in
large density fluctuations in the early universe, with masses ranging from one
Planck mass to a few solar masses, which is the mass within the horizon at
the QCD phase transition (phase transitions usually produce fluctuations
large enough to cause the collapse of the whole causally connected region).
Electroweak densities are reached very easily in such violent processes
where usual homologous approximation clearly fails. In that case, however,
the electroweak stars should be very long lived in order to be of astrophysical significance.
One more caveat is of course that so far no primordial black holes have been observed.

Studying the details of the the electroweak star physics, may give some clues for understanding of the supernova explosion mechanism and phenomenology on the side track. Also, the electroweak processes give an effective mechanism of erasing baryon number from a collapsing star before it becomes a black hole, which may be interesting from the point of view of the information loss paradox (see discussion in \cite{Stojkovic:2005zq}).

Seeing the gravitational collapse as an inverse Big Bang process, one might ask what happens when the next phase transition is hit, i.e the Grand Unifying Theory (GUT) phase transition. Our preliminary estimates indicate that the core of such an object should be microscopic. It is very unlikely that this stage will be reached before the object crosses its own Schwarzschild radius.

\acknowledgments
DS acknowledges the financial support from NSF, grant number PHY-0914893.
 GDS and AL thanks CERN for its hospitality.   GDS is supported by a grant from the US DOE to the
particle astrophysics theory group at CWRU.

\end{document}